\documentclass[12pt,a4paper]{article}
\usepackage[latin1]{inputenc}
\usepackage{amsmath}
\usepackage{appendix}
\usepackage{multirow}
\usepackage{url}
\usepackage{tikz}
\usepackage{graphicx}

\def\bc{\begin{center}}
\def\ec{\end{center}}
\usepackage{amsfonts}
\date{}
\usepackage{amsthm}
\usepackage{amssymb}
\usepackage{graphicx} \graphicspath{ {seipati/} }
\usepackage{amsmath}
\usepackage{color}
\usepackage{fancyhdr}
\usepackage{verbatim}
\usepackage{amsfonts,longtable}
\usepackage{epsfig}
\usepackage{amsfonts}
\usepackage{color}

\def\R{\hbox{{\rm I}\kern-0.2em{\rm R}\kern0.2em}}%mathematical R for reals

\def\bn{\begin{equation}}
\def\en{\end{equation}}
\def\bny{\begin{eqnarray}}
\def\eny{\end{eqnarray}}
\def\be{\begin{eqnarray*}}
\def\ee{\end{eqnarray*}}
\def\bc{\begin{center}}
\def\ec{\end{center}}

\def\({\left(}
\def\){\right  )}
\def\[{\left[}
\def\]{\right]}
\def\bc{\begin{center}}
\def\ec{\end{center}}

\newtheorem{dfn}{Definition}[section]
\newtheorem{thm}{Theorem}[section]
\newtheorem{rem}{Remark}[section]
\newtheorem{pro}{Proposition}[section]

\newtheorem{cor}{Corollary}[section]
\newtheorem{lem}{Lemma}[section]
\newtheorem{exm}{Example}[section]

\def\bn{\begin{equation}}
\def\en{\end{equation}}
\def\bny{\begin{eqnarray}}
\def\eny{\end{eqnarray}}
\def\be{\begin{eqnarray*}}
\def\ee{\end{eqnarray*}}
\def\bdn{\begin{dfn}}
\def\edn{\end{dfn}}
\def\btm{\begin{thm}}
\def\etm{\end{thm}}
\def\bpf{\begin{proof}}
\def\epf{\end{proof}}
\def\bpn{\begin{pro}}
\def\epn{\end{pro}}
\def\brk{\begin{rem}}
\def\erk{\end{rem}}
\def\bcy{\begin{cor}}
\def\ecy{\end{cor}}
\def\blm{\begin{lem}}\def\elm{\end{lem}}
\def\bex{\begin{exm}}
\def\eex{\end{exm}}

 \def\R{{\hat R}}

\begin{document}

{\bc {\Large \bf Neutron stars in general relativity and
scalar-tensor theory of gravity}\ec} {\bc {\bf Farrukh J.
Fattoyev}\ec} {Department of Physics, Manhattan College, Riverdale,
New York, NY, 10471, USA}

\bc {\bf  Abstract}\ec {The masses and radii of neutron stars are
discussed in general relativity and scalar-tensor theory of gravity
and the differences are compared with the current uncertainties
stemming from the nuclear equation of state in the relativistic
mean-field framework. It is shown that astrophysical and
gravitational waves observations of radii of neutron stars with
masses $M \lesssim 1.4 M_{\odot}$ constrain only the nuclear
equation of state, and in particular the density dependence of the
nuclear symmetry energy. Future observations of massive neutron
stars may constrain the coupling parameters of the scalar-tensor
theory provided that a general consensus on the dense nuclear matter
equation of state is reached.} \bc {\bf Keywords}: modified theories
of gravity; scalar-tensor gravity; compact objects; nuclear equation
of state; relativistic models; symmetry energy \ec

\newpage

\section{Introduction}
\label{Introduction}

Neutron stars are ideal astrophysical laboratories to probe the
nature of nuclear matter under extreme conditions of density and
isospin asymmetry\,\cite{Lattimer:2000nx, Lattimer:2006xb,
Lattimer:2012nd, Steiner:2004fi, Steiner:2010fz} as well as to test
fundamental theories of strong-field gravity\,\cite{Will:2014kxa,
Yunes:2013dva, Psaltis:2008bb}. While there has been significant
improvement in understanding properties of dense nuclear matter at
and near nuclear saturation density, $\rho_0 \simeq 2.5 \times
10^{14}$ g$/$cm$^{3}$, our knowledge of dense matter at
super-saturation densities $\rho > \rho_0$ corresponding to the core
region of neutron stars remains quite poor. In part, it is related
due to the fact that current nuclear interaction models that are
fitted to properties of terrestrial nuclear observables largely fail
to constrain the isovector part of the nuclear interaction. This, in
turn, affects model predictions for properties of nuclear matter
with large isospin asymmetry that are present in the core of neutron
stars\,\cite{Horowitz:2000xj, Horowitz:2001ya, Fattoyev:2012ch}. As
a result different neutron-star matter equation of state models
emerge that give rise to very different neutron-star structure
properties, such as masses and radii\,\cite{Fattoyev:2010mx,
Fattoyev:2010rx}, moments of inertia\,\cite{Fattoyev:2010tb,
Steiner:2014pda}, and tidal deformations\,\cite{Fattoyev:2012uu,
Fattoyev:2017jql}. Moreover, there is a possible degeneracy between
the nuclear matter equation of state and models of gravity applied
to describe the structure of neutron stars. While many studies have
been devoted to break this degeneracy, it remains one of the
outstanding problems to date\,\cite{He:2014yqa, Sotani:2017pfj,
Sotani:2004rq, Lasky:2008fs, Wen:2009av, Cooney:2009rr,
Arapoglu:2010rz, Pani:2011xm, Sotani:2012eb, Xu:2012wc, Lin:2013rea,
Sotani:2014goa, Eksi:2014wia}.

In this contribution, we examine effects of the nuclear equation of
state uncertainties simultaneously within the general theory of
relativity (GR) and the scalar-tensor (ST) theory of gravity. For
over a century, GR has been tested in many astrophysical scenarios
and thus far its agreement with experiments and observations have
been remarkable\,\cite{Sotani:2017pfj}. Nevertheless most of the
general relativistic tests have been performed in the weak-field
regime\,\cite{Psaltis:2008bb}. Neutron stars exhibit a strong
curvature, and it is not yet clear whether gravitational field of
such compact objects is fully described by GR. Astrophysical
observations may help pin down theories of gravitation in the
strong-field regime provided that the equation of state of dense
nuclear matter is well-known at all densities relevant for neutron
stars. In this article, we show that future astrophysical and
gravitational wave observations from low- and canonical-mass neutron
star will constrain the equation of state of nuclear matter, whereas
measurement of radii and tidal deformations of massive neutron stars
will aid in putting further constraints on the ST theories of
gravity.

In Sec. II we present the formalism of calculating the structure of
neutron stars in the scalar-tensor theory of gravity. Next, in Sec.
III, we review the equation of state of neutron-star matter within
the relativistic mean-field framework and discuss, in particular,
the current uncertainties in the isovector sector of the nuclear
interaction. In Sec. IV, we present our results for the neutron-star
structure calculation in both GR and ST theory of gravity. We
demonstrate that the equation of state of nuclear matter can be
constrained using canonical- and low-mass neutron star observations
irrespective of the models of gravity used. We also show that once
the equation of state is constrained, in conjunction with the
massive neutron star observations, one may then probe and limit the
parameter space of the ST theory of gravity.

Throughout this paper, we adopt geometric units, $c = 1 = G$, where
$c$ is the speed of light, and $G$ is the gravitational constant,
respectively.

\section{Neutron Star Structure in Scalar-Tensor Theory of Gravity}
\label{ScalarTensor}

The scalar-tensor theories of gravitation are one of the most
natural generalizations of general relativity that dates back to
early 1950s\,\cite{Jordan:1949zz, Fierz:1956zz, Jordan:1959eg,
Brans:1961sx, Dicke:1961gz}. According to these theories, the
gravitational force is also mediated by a scalar field, $\varphi$,
in addition to the second-rank metric tensor, $g_{\mu\nu}$, present
in general relativity. One can write the most general form of the
action defining this theory
as~\cite{Damour:1996ke,Yazadjiev:2014cza}
\begin{equation}
 S = \frac{1}{16 \pi} \int d^4x
\sqrt{-g_{\ast}}\left[R_{\ast} -2 g_{\ast}^{\mu \nu} \varphi_{,\mu}
\varphi_{, \nu} - V(\varphi)\right] + S_{\rm matter}\left[\psi_{\rm
matter}; A^2(\varphi)g_{\ast \mu\nu} \right] \ ,
\end{equation}
where $R^{\ast} \equiv g_{\ast}^{\mu \nu} R^{\ast}_{\mu \nu}$ is the
curvature scalar of the so-called ``Einstein metric", $g_{\ast
\mu\nu}$, describing the pure spin-2 excitations, whereas $\varphi$
is a long-range scalar field describing spin-0 excitations. Here
$V(\varphi)$ is the scalar field potential, $\psi_{\rm matter}$ is a
collective representation of all matter fields, and $S_{\rm matter}$
is the corresponding action of the matter represented by $\psi_{\rm
matter}$, which in turn is coupled to the ``Jordan-Fierz metric''
$g_{\mu\nu}$ that is related to the ``Einstein metric" through the
conformal transformation:
\begin{equation}
g_{\mu\nu} \equiv A^2(\varphi)g_{\ast \mu\nu} \ .
\end{equation}
The field equations are easily formulated in the ``Einstein metric",
however all non-gravitational physical experiments measure
quantities in the ``Jordan-Fierz metric" and thus they are referred
to as the ``physical metric". From here on, we refer to them as the
Einstein frame and the Jordan frame, respectively.

Taking variation of the action $S$ with respect to the metric
$g_{\ast \mu\nu}$ and the scalar field $\varphi$, we find the set of
field equations in the Einstein frame
\begin{eqnarray}
&& G_{\ast \mu \nu} = 8 \pi T_{\ast \mu \nu} +  2 \varphi_{,\mu}
\varphi_{, \nu} - g_{\ast \mu \nu} g_{\ast}^{\alpha \beta}
\varphi_{,\alpha} \varphi_{, \beta} - \frac{1}{2} V(\varphi) g_{\ast \mu \nu} \ , \\
&& \Box_{\ast} \varphi - \frac{1}{4} \frac{d V(\varphi)}{d\varphi} =
- 4 \pi \alpha(\varphi) T_{\ast} \ .
\end{eqnarray}
Here again $T_{\ast \mu \nu}$ is the energy-momentum tensor in the
Einstein frame, which is related to the physical energy-momentum
tensor in Jordan frame $T_{\mu \nu}$ via
\begin{equation}
T_{\ast \mu \nu} = A^2(\varphi) T_{\mu \nu}
\end{equation}
and $T_{\ast}$ and $\alpha(\varphi)$ are defined as
\begin{eqnarray}
&& T_{\ast} \equiv T_{\ast \mu}^{\mu} \ , \\
&& \alpha(\varphi) \equiv \frac{d \ln A(\varphi)}{d \varphi} \ .
\end{eqnarray}
The energy-momentum conservation can either be expressed as
$\nabla_{\nu} T_{\mu}^{\nu} = 0$ in the Jordan frame, or
\begin{equation}
\nabla_{\ast \nu} T_{\ast \mu}^{\nu} = \alpha(\varphi) T_{\ast}
\nabla_{\ast \mu} \varphi \ ,
\end{equation}
in the Einstein frame.

We consider the neutron star to be made of a perfect fluid. In this
case the energy density $\mathcal{E}$, the pressure $P$ and the
4-velocity $u_{\mu}$ in the two frames are related via the following
relations
\begin{eqnarray}
&& \mathcal{E}_{\ast} = A^4(\varphi) \mathcal{E} \ , \\
&& P_{\ast} = A^4(\varphi) P \ , \\
&& u_{\ast \mu} = A^{-1}(\varphi) u_{\mu} \ ,
\end{eqnarray}
where the subscript asterisks denote quantities in the Einstein
frame, as usual. The spacetime metric describing an unperturbed,
non-rotating, spherically symmetric neutron star can be written as
\begin{equation}
ds^2_{\ast} = -e^{2 \nu(r)} dt^2 + e^{2 \lambda(r)} r^2 + r^2
(d\theta^2 + \sin^2 d\phi^2) \ ,
\end{equation}
where
\begin{equation}
e^{-2 \lambda(r)} = 1 - \frac{2 M(r)}{r} \equiv N^2(r) \ ,
\end{equation}
and the function $\nu(r)$ will be calculated later. Here we also
defined $N(r) \equiv \left(1-2M(r)/r \right)^{1/2}$, for
convenience. Using the space-time metric above, one can arrive at
the dimensionally reduced field equations\,\cite{Yazadjiev:2014cza},
which in turn can be cast into equations for hydrostatic structure:
\begin{eqnarray}
\label{hydrostaticSTa} \frac{dM(r)}{dr} &=& 4 \pi r^2  A^4(\varphi)
\mathcal{E}(r) + \frac{r^2}{2} N^2(r) \chi^2(r) +
\frac{r^2}{4} V(\varphi) \ , \\
\label{hydrostaticSTb} \frac{dP(r)}{dr} &=& -\bigg(\mathcal{E}(r) +
P(r)\bigg) \bigg(\frac{d \nu(r)}{dr} + \alpha(\varphi) \chi(r)\bigg)
\ ,
\\
\label{hydrostaticSTc} \frac{d\nu(r)}{dr} &=& \frac{M(r)}{r^2
N^2(r)} + \frac{4 \pi r}{N^2(r)} A^4(\varphi) P(r)  +  \frac{1}{2} r
\chi^2(r) -\frac{r}{4 N^2(r)} V(\varphi) \ ,
\\
\nonumber \frac{d \chi(r)}{dr} &=&  \bigg\{4 \pi A^4(\varphi)
\bigg[\alpha(\varphi) \bigg(\mathcal{E}(r) - 3 P(r)\bigg) +
r \chi(r)\bigg(\mathcal{E}(r) - P(r)\bigg)\bigg] -   \ \\
\label{hydrostaticSTd} &-& \frac{2 \chi(r)}{r}
\left(1-\frac{M(r)}{r}\right) + \frac{1}{2}r\chi(r)V(\varphi) +
\frac{1}{4}\frac{dV(\varphi)}{d\varphi}  \bigg\}N^{-2}(r)  \ .
\end{eqnarray}
where we introduced
\begin{equation}
\chi(r) \equiv \frac{d \varphi}{dr} \ .
\end{equation}
It is well known that predictions of scalar-tensor theories are
physically equivalent to non-linear modified gravity
theories\,\cite{Magnano:1993bd}. In particular, for the
$R^2$-gravity, where the Ricci scalar is replaced with $f(R) = R + a
R^2$ in the Einstein-Hilbert action, the explicit form of the
potential $V(\varphi)$ can be written as
\begin{equation}
V(\varphi) = \frac{1}{4a} \left(1-e^{-2\varphi}{\sqrt{3}}\right)^2 \
, \qquad \alpha(\varphi) = -\frac{1}{\sqrt{3}} \ .
\end{equation}

Next, following Ref.~\cite{Damour:1996ke} we set $V(\varphi)=0$, and
consider a coupling function of the form
\begin{equation}
A(\varphi) = \exp\left(\alpha_0\varphi + \frac{1}{2}\beta_0
\varphi^2\right) \ .
\end{equation}
The coupling constants $\alpha_0$ and $\beta_0$ are real numbers. It
was shown that measurement of the surface atomic line redshifts from
neutron stars could be used as a direct test of strong-field gravity
theories\,\cite{DeDeo:2003ju}. In particular, coupling constants
with $\alpha_0=0$ and $\beta_0=-8$ were used. Over the past decade,
significant improvements have been made in constraining these
coupling constants. For example, solar-system experiments,
binary-pulsar and pulsar-white dwarf timing observations put some of
the most stringent constraints on these constants that
conservatively can be written as\,\cite{Damour:1996ke,
Freire:2012mg}
\begin{eqnarray}
&& \alpha_0^2 = \partial \ln A(\varphi_0) / \partial \varphi_0
\lesssim 2 \times
10^{-5} \ , \\
&& \beta_0 = \partial^2 \ln A(\varphi_0) / \partial \varphi_0^2
\gtrsim -5 \ .
\end{eqnarray}
On the other hand, it was first shown in Ref.~\cite{Damour:1996ke}
that predictions by models with $\beta_0>-4.35$ can not in general
be distinguished from the general relativistic results due to the
so-called ``spontaneous scalarization" effect. Notice that GR is
automatically recovered in the limits of $\alpha_0 = 0$ and $\beta_0
= 0$, where the equations
(\ref{hydrostaticSTa}--\ref{hydrostaticSTd}) are reduced to the
famous Tolman Oppenheimer Volkoff equations\,\cite{Tol39_PR55,
Opp39_PR55}.

We solve the interior and the exterior problem simultaneously using
the following natural boundary conditions in the center of the star
\begin{equation}
P(0) = P_{\rm c} \ , \qquad \mathcal{E}(0) = \mathcal{E}(\rm c) \ ,
\qquad \chi(0) = 0 \ .
\end{equation}
We also demand cosmologically flat solution at infinity to agree
with the observation:
\begin{equation}
\lim_{r \rightarrow \infty} \nu(r) = 0 \ , \qquad \lim_{r
\rightarrow \infty} \varphi(r) = 0 \ .
\end{equation}
The only input required to integrate the Eqns.
(\ref{hydrostaticSTa}--\ref{hydrostaticSTd}) is the equation of
state of neutron-star matter, $\mathcal{E} = \mathcal{E}(P)$, that
is trivial in the case of exterior solution. Once an EOS is
supplemented, for a given central pressure $P(0) = P_{\rm c}$, one
can integrate them from the center of the star $r = 0$, all the way
up to $r \rightarrow \infty$.

The stellar coordinate radius is then determined by the condition in
which the pressure vanishes, \emph{i.e.} $P(r_{\rm s}) = 0$, where
$r_{\rm s}$ is the surface radius in the Einstein frame. The
physical radius of a neutron star $R$ is then found in the Jordan
frame through
\begin{equation}
R  = A\left(\varphi(r_{\rm s})\right)r_{\rm s} \ .
\end{equation}
The physical stellar mass $M$ as measured by an observer at
infinity---also known as the Arnowitt-Deser-Misner (ADM)
mass---matches with the coordinate mass, since at infinity the
coupling function approaches unity.

\section{Neutron-Star Matter Equation of State}
\label{EOS}

The structure of neutron stars is sensitive to the equation of state
of cold, fully catalyzed, and neutron-rich matter. The matter inside
neutron stars span many orders of magnitude in density leading to
rich and exotic phases in their interiors. In the outer crust of
neutron stars the matter is organized into a Coulomb lattice of
neutron-rich nuclei embedded in a degenerate electron
gas~\cite{Baym:1971pw,RocaMaza:2008ja}. The nuclear composition in
this region is solely determined by the masses of neutron-rich
nuclei in the region of $26 <  Z \lesssim 40$ and the pressure
support is primarily provided by the degenerate electrons. The
equation of state for this region is therefore relatively well
known\,\cite{Baym:1971pw, Haensel:1981aa}. In this work, we adopt
the outer crust equation of state by Haensel, Zdunik and
Dobaczewski\,\cite{Haensel:1981aa}. As the density increases, the
nuclei in the outer crust become more and more neutron-rich. At a
density of about $\rho \approx 4 \times 10^{11}$ g$/$cm$^{-3}$ the
nuclei in the outer crust become so neutron-rich that they can no
longer hold additional neutrons, and neutrons start dripping out.
This region defines the boundary between the outer and the inner
crust.

The inner crust extends from the neutron-drip density up to about
$\rho \approx 2/3 \rho_0$ where the uniformity in the system is
restored. On the top layers of the inner crust nucleons continue to
cluster into a Coulomb crystal of neutron-rich nuclei embedded in a
uniform electron gas, where however now the system is also in a
chemical equilibrium with a superfluid neutron
gas\,\cite{Piekarewicz:2013dka}. As the density continues to
increase, the spherical nuclei start to deform in an effort to
reduce the Coulomb repulsion. As a result, the inner crust exhibit
complex and exotic structures that are collectively known as
``nuclear
pasta"~\cite{Ravenhall:1983uh,Hashimoto:1984,Lorenz:1992zz}, which
emerge from a dynamical competition between the short-range nuclear
attraction and the long-range Coulomb repulsion. Although
significant progress has been made in simulating this exotic
region~\cite{Horowitz:2004yf,Horowitz:2004pv,Horowitz:2005zb,
Fattoyev:2017zhb}, the equation of state for this region remains
highly uncertain and must be inferred from theoretical calculations.
While a detailed knowledge of the equation of state for this region
is important for the interpretations of cooling
observations\,\cite{Fattoyev:2017ybi}, its impact on the bulk
properties of neutron stars is minimal\,\cite{Piekarewicz:2014lba}.
For this region, therefore we resort to the equation of state
provided by Negele and Vautherin\,\cite{Negele:1971vb}.

The structure of neutron stars are mostly sensitive to the equation
of the state of the core, and the crust-core transition properties.
In particular, most of the mass of neutron stars are contained in
the liquid core that comprises a region of the star with densities
of as low as one-third to as high as ten times nuclear-matter
saturation density $\rho_0$. For this region, we employ the equation
of state generated from various refinements of relativistic
mean-field model by Serot and
Walecka~\cite{Walecka:1974qa,Serot:1984ey,Serot:1997xg}. For
consistency, the transition density from the liquid core to the
solid crust is computed using the same relativistic mean-field
models, where it is done by searching for the critical density at
which the uniform system becomes unstable to small amplitude density
oscillations~\cite{Carriere:2002bx}. We would like to emphasize that
the crust-core transition density (hence transition pressure) plays
an important role in neutron star bulk properties with various
models leading to crust thicknesses that differ by over one km and
predicting significantly different crustal components of the moment
of inertia that are important in interpreting observations of pulsar
glitches\,\cite{Piekarewicz:2014lba}.

The equation of state for the uniform liquid core is based on an
interaction Lagrangian that has been accurately calibrated to a
variety of ground-state properties of both finite nuclei and
infinite nuclear matter. This model includes a nucleon field
($\psi$), a scalar-isoscalar meson field ($\phi$), a
vector-isoscalar meson field ($V^{\mu}$), and an isovector meson
field ($b^{\mu}$)~\cite{Serot:1984ey,Serot:1997xg}. The free
Lagrangian density for this model is given
by\,\cite{Fattoyev:2010tb, Fattoyev:2010mx} %%
\begin{eqnarray}
\nonumber {\mathcal{L}}_{0} &=& \bar\psi
\left(i\gamma^{\mu}\partial_{\mu}\!-\!m_b\right)\psi +
\frac{1}{2}\partial_{\mu} \phi \partial^{\mu} \phi
-\frac{1}{2}m_{s}^{2}\phi^{2} - \ \\
&-& \frac{1}{4}F^{\mu\nu}F_{\mu\nu} +
\frac{1}{2}m_{v}^{2}V^{\mu}V_{\mu} - \frac{1}{4}{\bf b}^{\mu\nu}{\bf
b}_{\mu\nu} + \frac{1}{2}m_{\rho}^{2}{\bf b}^{\mu}{\bf b}_{\mu} \;,
\label{Lagrangian0}
\end{eqnarray}
where $F_{\mu\nu}$ and ${\bf b}_{\mu\nu}$ are the nuclear isoscalar
and isovector field tensors, respectively%%
\begin{eqnarray}
 F_{\mu\nu} &=& \partial_{\mu}V_{\nu} - \partial_{\nu}V_{\mu} \;, \\
 {\bf b}_{\mu\nu} &=& \partial_{\mu}{\bf b}_{\nu}
 - \partial_{\nu}{\bf b}_{\mu} \;.
\label{FieldTensors}
\end{eqnarray}
Here, the parameters $m_b$, $m_{s}$, $m_{v}$, and $m_{\rho}$
represent the nucleon and meson masses and may be treated as
empirical constants. The interacting component of Lagrangian density
can be written by the following expression~\cite{Serot:1984ey,
Serot:1997xg,Mueller:1996pm} %%
\begin{equation}
{\mathcal{L}}_{\rm int} = \bar\psi \left[g_{\rm s}\phi   \!-\!
         \left(g_{\rm v}V_\mu  \!+\!
    \frac{g_{\rho}}{2}\mbox{\boldmath$\tau$}\cdot{\bf b}_{\mu}
          \right)\gamma^{\mu} \right]\psi -
          U(\phi,V^{\mu},{\bf b^{\mu}}) \ ,
\label{Lagrangian}
\end{equation}
that includes Yukawa couplings---with coupling parameters, $g_{\rm
s}$ , $g_{\rm v}$, and $g_{\rho}$---between the nucleon and meson
fields. The Lagrangian density is also supplemented by nonlinear
meson interactions, $U(\phi,V^{\mu},{\bf b}^{\mu})$ that improve the
phenomenological standing of the model,
%%%
\begin{equation}
U(\phi,V^{\mu},{\bf b}^{\mu}) =
   \frac{\kappa}{3!} (g_{\rm s}\phi)^3 \!+\!
    \frac{\lambda}{4!}(g_{\rm s}\phi)^4
 \!-\! \frac{\zeta}{4!}
    \Big(g_{\rm v}^2 V_{\mu}V^\mu\Big)^2 \!-\!
    \Lambda_{\rm v}
    \Big(g_{\rho}^{2}\,{\bf b}_{\mu}\cdot{\bf b}^{\mu}\Big)
    \Big(g_{\rm v}^2V_{\nu}V^\nu\Big) \;.
\label{USelf}
\end{equation}
%%%
The details on the calibration procedure can be found in
Refs.\,\cite{Serot:1984ey,Serot:1997xg,Horowitz:2000xj,Todd:2003xs,
Chen:2014sca} and references therein.

While the full complexity of the quantum system can not be tackled
exactly, the ground-state properties of the system may be computed
in a {\sl mean-field} approximation. In this approximation, all the
meson fields are replaced by their classical expectation values and
their solution can be readily obtained by solving the classical
Euler-Lagrange equations of motion. The only remnant of quantum
behavior is in the treatment of the nucleon field which emerges from
a solution to the Dirac equation in the presence of appropriate
scalar and vector potentials~\cite{Serot:1984ey,Serot:1997xg}.
Following standard mean-field practices, the energy density of the
system is given by the following expression: %%
\begin{eqnarray}
{\mathcal E} (\rho, \alpha) &=&
 \frac{1}{\pi^{2}}\int_{0}^{k_{\rm F}^{p}} k^{2}E_{k}^{\ast}\,dk +
 \frac{1}{\pi^{2}}\int_{0}^{k_{\rm F}^{n}} k^{2}E_{k}^{\ast}\,dk +
   \frac{1}{2}m_{s}^{2}\phi_{0}^{2} +
   \frac{\kappa}{3!} (g_{\rm s}\phi_{0})^3 +
   \frac{\lambda}{4!}(g_{\rm s}\phi_{0})^4 +
 \nonumber \\
  &+&
   \frac{1}{2}m_{v}^{2}V_{0}^{2} +
   \frac{\zeta}{8}(g_{\rm v}V_{0})^4 +
   \frac{1}{2}m_{\rho}^{2}b_{0}^{2} +
   3\Lambda_{\rm v}(g_{\rm v}V_{0})^2 (g_{\rho}b_{0})^2 \;.
\label{EDensity}
\end{eqnarray}
where $\rho$ is the baryon density of the system, $\alpha = (
\rho_{\rm n} - \rho_{\rm p})/\rho$ is the neutron-proton asymmetry,
$E_{k}^{\ast}\!=\!\sqrt{k^{2}+m_{\rm b}^{\ast 2}}$, $m_{\rm
b}^{\ast}\!=\!m_{\rm b}-g_{\rm s}\phi_{0}$ is the effective nucleon
mass, $k_{\rm F}^{p} (k_{\rm F}^{n})$ is the proton (neutron) Fermi
momentum. Since the mean-field approximation is thermodynamically
consistent, the pressure of the system at zero temperature may be
obtained either directly from the energy-momentum tensor or from the
energy density and its first
derivative~\cite{Serot:1984ey,Serot:1997xg}. That is, %%
\begin{equation}
 P (\rho, \alpha) = \rho \frac{\partial \mathcal{E}(\rho, \alpha)}{\partial \rho }
 -\mathcal{E}(\rho, \alpha) \;, \label{Pressure}
\end{equation}
It is often useful to expand the energy per nucleon of the system in
even powers of $\alpha$:
\begin{equation}
E/A(\rho, \alpha) - m_b = E_{\rm SNM}(\rho) + \alpha^2 S(\rho) +
\mathcal{O}(\alpha^4) \ ,
\end{equation}
where $E_{\rm SNM}(\rho)$ is the energy per nucleon of symmetric
nuclear matter, whereas $S(\rho)$ is referred to as the nuclear
symmetry energy, a quantity that represents the increase in the
energy of the system as it departs from the symmetric limit of equal
number of neutrons and
protons\,\cite{Tsang:2012se,Horowitz:2014bja}.

Further, it is customary to characterize the behavior of both
symmetric nuclear matter and the symmetry energy in terms of a few
bulk parameters near nuclear saturation density $\rho_0$:
\begin{eqnarray}
&& E_{\rm SNM}(\rho) = \varepsilon_0 + \frac{1}{2} K x^2 + \ldots \ , \\
&& S(\rho) = J + L x + \frac{1}{2} K_{\rm sym} x^2 + \ldots \ ,
\end{eqnarray}
where $x \equiv (\rho -\rho_0)/3\rho_0$ is a dimensionless
parameter, $\varepsilon_0$ and $K$ represent the energy per nucleon
and the incompressibility coefficient of symmetric nuclear matter,
respectively, whereas $J$, $L$, and $K_{\rm sym}$ are the magnitude,
slope and curvature of the symmetry energy at saturation. The bulk
parameters of symmetric nuclear matter are relatively
well-constrained\,\cite{Fattoyev:2012rm, Fattoyev:2012uu,
Fattoyev:2012rm}. On the other hand, the density dependence of the
nuclear symmetry energy remains unconstrained due to lack of
sensitive isovector nuclear probes\,\cite{Fattoyev:2013yaa,
Li:2013ola}. In this contribution, we will primarily concentrate on
the impact of variations of density slope of the symmetry energy $L$
that is closely related to the pressure of pure neutron matter at
saturation density.

We assume the neutron-star matter to consist of neutrons, protons,
electrons, and muons in chemical equilibrium. We do not consider any
``exotic" degrees of freedom, such as hyperons, meson condensates,
or quarks. The electrons and muons are assumed to behave as
relativistic free Fermi gases (with $m_{e}\!\equiv\!0$). The muons
appear in the system only after the electronic Fermi momentum
becomes equal to the muon rest mass. The total energy density and
pressure of the star are obtained by adding up the nucleonic and
leptonic contributions.

\section{Results}
\label{Results}

In this work, we use the FSUGold2 parametrization that was
introduced by Ref.\,\cite{Chen:2014sca} to specifically apply for
both finite nuclei and neutron stars. Of particular importance to
this study is the role of omega-meson self-interactions, as
described by the parameter $\zeta$ in the interaction Lagrangian
density, that is used to tune the equation of state at high density
to reproduce the maximum mass of a neutron star. It was first shown
by M\"uller and Serot that by using different values of $\zeta$ one
can reproduce the same observed nuclear matter properties at nuclear
saturation, yet produce maximum neutron star masses---using general
relativity only---that differ by almost one solar
mass~\cite{Mueller:1996pm}. For example, models with $\zeta\!=\!0$
predict the maximum neutron star masses of about $2.8 M_{\odot}$.
Note that this tuning primarily affects the equation of state of
symmetric nuclear matter at high density, which is relevant to the
core of neutron stars. On the other hand, by including the nonlinear
coupling constant $\Lambda_{\rm v}$, Horowitz and Piekarewicz showed
that one can modify the density-dependence of the symmetry
energy\,\cite{Horowitz:2000xj}. It was shown that tuning
$\Lambda_{\rm v}$ provides a simple and efficient method of
controlling the density dependence of symmetry energy without
compromising the success of the model in reproducing well determined
ground-state observables. The original FSUGold2 model has a
relatively stiff symmetry energy with the density slope of $L =
112.8$ MeV. Following the same method as outlined in
Ref.\,\cite{Fattoyev:2010tb}, we obtain a family of ``FSUGold2"
parametrizations with $L = 47$, $60$, $80$, $100$ MeV. We emphasize
that in doing so predictions for properties of finite nuclei, such
as the binding energies and charge radii of closed shell nuclei
remain intact.

The maximum mass of a neutron star predicted by the original
FSUGold2 with $\zeta = 0.0256$ and using the general relativistic
TOV equations is $M_{\rm max} = 2.07 M_{\odot}$, which is consistent
with the observations of highly precise measurements of two massive
neutron stars made at the Green Bank
Telescope\,\cite{Demorest:2010bx, Antoniadis:2013pzd}. Indeed, the
maximum possible mass of a neutron star may not be very far from
this value as was recently shown by Refs.\,\cite{Margalit:2017dij,
Rezzolla:2017aly,Most:2018hfd}, $M_{\rm max} \approx 2.17 M_{\rm
Sun}$. This in turn suggests that the $\zeta$ parameter that
controls the stiffness of the equation of state of symmetric nuclear
matter is already well-constrained, and future observations of
maximum mass may put even tighter constraints.

On the other hand, the original FSUGold2 model predicts the
corresponding general relativistic radius of a canonical 1.4
solar-mass neutron star to be $R =14.11$ km. Unfortunately, direct
determination of the neutron star radii at present is not quite
satisfactory. Early attempts by \"{O}zel and collaborators to
determine simultaneously the mass and radius of three x-ray bursters
resulted in stellar radii to be between $8$ and $10$
km\,\cite{Ozel:2010fw}. Later, Steiner {\sl{et al.}} supplemented
\"{O}zel's study with additional neutron stars and concluded that
the most probable radius of a 1.4 $M_{\odot}$ lie in the range of
10.4-12.9 km\,\cite{Steiner:2010fz, Steiner:2012xt}. Nevertheless,
this more conservative estimate has been put into question by
Suleimanov {\sl{et al.}}, who used a more complete model of the
neutron star atmosphere and obtained a radius greater than 14 km for
a single source studied\,\cite{Suleimanov:2010th}. Recognizing this
situation and the many challenges posed by the study of x-ray
bursters, Guillot {\sl{et al.}} concentrated on the determination of
stellar radii by studying quiescent low mass x-ray binaries (qLMXB)
in globular clusters. By explicitly stating all their assumptions,
in particular the common radius assumption, they were able to
determine a rather small neutron star radius of $9.4 \pm 1.2$
km\,\cite{Guillot:2013wu, Guillot:2014lla}. And recently, an
improved precision over previous measurements was obtained by
incorporating distance uncertainties to the globular cluster M13
that suggests the neutron star radius to be $12.3^{+1.9}_{-1.7}$ km
and/or $15.3^{+2.4}_{-2.2}$ km depending on the composition of the
atmosphere being as H or He, respectively\,\cite{Shaw:2018wxh}.

%%%%%%%%%%%%%%%%%%%%%%%%%%%%%%%%%%%%%%%%%%%%%%%%%%%%%%%%%%%%%%%%%%%%
% START: This paragraph was added/modified in response to
% Referee's remark.
%%%%%%%%%%%%%%%%%%%%%%%%%%%%%%%%%%%%%%%%%%%%%%%%%%%%%%%%%%%%%%%%%%%%

Since the density slope of the symmetry energy $L$ is related to the
pressure of pure neutron matter at saturation density, and the
pressure of the neutron-rich matter is strongly correlated with the
neutron star radius\,\cite{Brown:2000pd, Fattoyev:2012rm}, we build
a family of the FSUGold2 parametrizations by using different values
of $L$. While these parametrizations provide an accurate description
of ground-state properties of finite nuclei, they also predict
stellar radii that differ by over one km. The uncertainties in the
density dependence of the symmetry energy exhibited by the RMF
models here broadly bracket the uncertainties stemming from various
nuclear many-body models. This include from non-relativistic
mean-field Skyrme interactions to microscopic calculations of the
equation of state that are based on the nucleon-nucleon interactions
and consider other degrees of freedom such as pions,
$\Delta$-resonances, and hyperons\,\cite{Fattoyev:2012ch,
Steiner:2011ft, Lonardoni:2014bwa,Cai:2015hya}. Note however that
none of our parametrizations can produce neutron star radii that are
smaller than $12$ km, which remains one of the biggest challenges
today\,\cite{Fattoyev:2010rx, Jiang:2015bea}.
%%%%%%%%%%%%%%%%%%%%%%%%%%%%%%%%%%%%%%%%%%%%%%%%%%%%%%%%%%%%%%%%%%%%
% END: This paragraph was added/modified in response to
% Referee's remark.
%%%%%%%%%%%%%%%%%%%%%%%%%%%%%%%%%%%%%%%%%%%%%%%%%%%%%%%%%%%%%%%%%%%%

Fortunately, the prospects for precision neutron star mass-radius
measurements have never been better, especially with the upcoming
Neutron Star Interior Composition Explorer (NICER) x-ray timing
mission that can put some of the tightest constraints in the near
future. On the other hand, the recent gravitational wave observation
from a binary neutron star merger\,\cite{TheLIGOScientific:2017qsa}
has already put some indirect constraint on the neutron star radii.
In particular, using the upper limit on the \emph{tidal
deformability} measurement determined by LIGO and Virgo
Collaboration, Refs.\,\cite{Fattoyev:2017jql,Annala:2017llu,
Tews:2018chv} have significantly constrained the allowed EoS models,
which in turn predicted a radius of a canonical neutron star to be
smaller than about 14 km. The tidal deformability is an intrinsic
neutron-star property that describes the tendency of a neutron star
to develop a mass quadrupole as a response to the tidal field
induced by its companion\,\cite{Damour:1991yw, Flanagan:2007ix}. The
dimensionless tidal polarizability $\Lambda$ is defined as follows:
%%%
\begin{equation}
 \Lambda = \frac{2k_{2}}{3}\left(\frac{R}{M}\right)^{5} \;,
 \label{Lambda}
\end{equation}
%%%
where $k_{2}$ is the second Love number\,\cite{Binnington:2009bb,
Damour:2012yf}. The tidal deformability is highly sensitive to the
stellar radius, $\Lambda\!\sim\!\!R^{5}$, and therefore its
measurement can be used as a proxy to constrain the neutron star
radius that has been notoriously difficult to measure in the
past\,\cite{Ozel:2010fw, Steiner:2010fz, Suleimanov:2010th,
Guillot:2013wu, Lattimer:2013hma, Heinke:2014xaa, Guillot:2014lla,
Ozel:2015fia, Steiner:2017vmg, Nattila:2017wtj}.
%\begin{widetext}
\begin{table}[t]
\scalebox{0.67}{
\begin{tabular}{|c|c|c|c|c|c|c|c|c|}
\hline $L$ (MeV) & $R_{1.4}^{\rm GR}$ (km) & $R_{1.4}^{\rm ST}$ (km)
& $R_{2.0}^{\rm GR}$ (km) & $R_{2.0}^{\rm ST}$ (km) & $M_{\rm
max}^{\rm GR}$ ($M_{\odot}$) & $M_{\rm max}^{\rm ST}$ ($M_{\odot}$)
& $R_{\rm max}^{\rm GR}$ (km) & $R_{\rm max}^{\rm ST}$ (km) \\ %[0.5ex]
\hline
~47.0 & 12.77 & 12.75 & 11.99 & 12.66 & 2.027 & 2.112 & 11.55 & 12.20  \\
~60.0 & 13.08 & 13.06 & 12.09 & 12.81 & 2.024 & 2.109 & 11.64 & 12.33  \\
~80.0 & 13.42 & 13.41 & 12.25 & 12.97 & 2.025 & 2.110 & 11.78 & 12.46  \\
100.0 & 13.81 & 13.80 & 12.59 & 13.21 & 2.041 & 2.128 & 11.97 & 12.65  \\
112.8 & 14.11 & 14.11 & 12.96 & 13.46 & 2.073 & 2.159 & 12.14 & 12.84  \\
\hline
\end{tabular}}
\caption{Predictions for the masses and radii of neutron stars in
both general relativity and scalar-tensor theory of gravity using a
family of FSUGold2 interaction whose isovector coupling constants
are tuned to give different values of the density slope of the
symmetry energy, $L$. Here superscripts ``GR" and ``ST" stand for
general relativity and scalar-tensor theory of gravity,
respectively, and subscripts ``$1.4$", ``$2.0$", and ``max" stand
for $1.4$-, $2.0$- and maximum mass stellar configurations.}
\label{Table1}
\end{table}
%\end{widetext}

%%%%%%%%%%%%%%%%%%%%%%%%%%%%%%%%%%%%%%%%%%%%%%%%%%%%%
\begin{figure}[h]
\smallskip
\includegraphics[width=3.5in,angle=0]{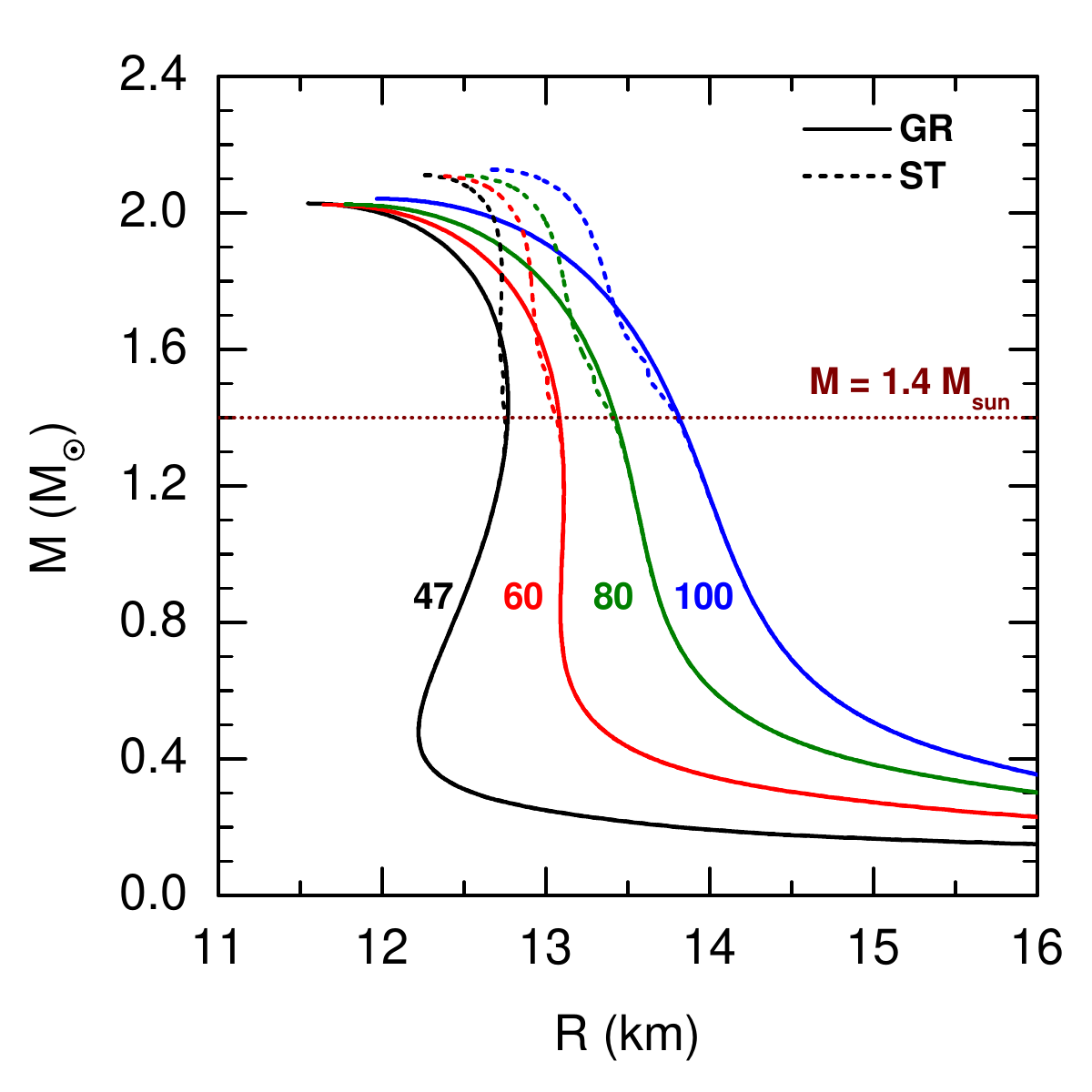}
\caption{The mass-versus-radius relation calculated using the family
of FSUGold2 parametrizations considered in this work labeled with
the corresponding $L$-values. Here $M$ stands for either general
relativistic TOV-mass (solid lines) or the ADM-mass (dashed lines)
predicted by the scalar-tensor theory. For the coupling constants of
scalar-tensor theory an upper observational bounds of $\alpha_0 =
\sqrt{2.0} \times 10^{-5}$ and $\beta_0 = -5.0$ are used.}
\label{Fig1}
\end{figure}
%%%%%%%%%%%%%%%%%%%%%%%%%%%%%%%%%%%%%%%%%%%%%%%%%%%%%

In Table \ref{Table1} we provide with our calculations for neutron
star masses and radii predicted by both general relativity and the
scalar tensor theory of gravity. Notice that the variations of the
density slope of the symmetry energy as depicted in various
$L$-values predict neutron star radii that are different by over one
kilometer. Nevertheless, both general relativity and scalar-tensor
theory predicts very similar stellar radii for most neutron stars,
unless their mass is close to two solar mass. This suggests that
measurements of canonical neutron-star radii, in particular, would
be unable to constrain the parameters of the scalar-tensor theory
but would be extremely useful in constraining the equation of state
of nuclear matter. Moreover, given that the density slope of the
symmetry energy is relatively insensitive to the maximum stellar
mass which mostly constrains the equation of state of symmetric
nuclear matter at high densities, one can use radii measurements to
place significant constraints primarily on the density dependence of
the nuclear symmetry energy. For massive neutron stars the radii
predictions in two models of gravity start to deviate. This
phenomenon of the so-called ``spontaneous scalarization" was first
observed by Damour and Esposito-Far\`{e}se\,\cite{Damour:1991yw}.

For completeness, in Fig.\,\ref{Fig1} we also show the full
mass-versus-radii relation predicted in both GR and ST. Since recent
measurements of tidal deformability suggested that neutron stars
with radii $R \gtrsim 14$ km may have already been ruled
out\,\cite{Fattoyev:2017jql, Annala:2017llu, Abbott:2018exr,
Most:2018hfd, Tews:2018chv, Malik:2018zcf, Tsang:2018kqj}, in this
figure we do not display predictions from the original FSUGold2
parametrization. It is safe to say that given that multi-messenger
era of gravitational wave astronomy and x-ray observations of
neutron stars is in its infancy, future observations of neutron
stars in mass-range of $M \lesssim 1.4 M_{\odot}$ will undoubtedly
put tighter constraints on the nuclear equation of state (below
$\rho \approx 2.5 \rho_0$ corresponding to the central density of a
neutron star) but not on the coupling constants of the scalar-tensor
theory.

It is particularly interesting to compare the radii of two-solar
mass neutron stars predicted by both theory of gravitation. As an
example, we use the soft equation of state with $L=47$ MeV and find
that stellar radii differ by 5.6\% (See Table I). Since massive
neutron stars are rarely found in nature---hence lots of theoretical
work and sufficient observational data may be required to
distinguish general relativity from the scalar-tensor theory using
radii observations from massive stars alone---we note that this may
not be the case if one considers tidal deformability measurements
which scales as $R^5$. In this contribution we did not directly
calculate tidal deformabilities in the scalar-tensor theory that in
addition to radii also depends on the tidal Love number
$k_2$\,\cite{Pani:2014jra}. It has been shown, however that within
the general relativistic framework $k_2$ is less sensitive to the
radius of a neutron star, and the $R^5$ scaling behavior in tidal
deformability is quite robust\,\cite{Fattoyev:2017jql}. It is worth
to point out that the ratio $(R^{\rm ST}_{2.0}/R^{GR}_{2.0})^5 =
1.312$ suggests that the corresponding tidal deformabilities in two
models may differ by over 30\%. We will explore this in more details
in future work. This is intriguing because while mass-radii
measurements of low mass neutron stars will constrain the equation
of state, tidal deformability measurements of massive stars would
enable to test models of gravity in strong regime.

\section{Conclusions}
\label{Conclusions}

We have examined bulk properties of neutron stars in general
relativity and the scalar-tensor theory of gravity. In particular,
we discussed the current uncertain role of the density dependence of
the nuclear symmetry energy in determination of the equation of
state of neutron star matter. Our analysis show that future
observations of neutron-star radii with masses $M \lesssim 1.4
M_{\odot}$ will enable to primarily constrain the equation of state
of dense nuclear matter but not the coupling constants of the
scalar-tensor theory of gravity. In particular, we have found that
in conjunction with the maximum stellar mass measurements, this will
lead to placing tight constraints on the density dependence of the
nuclear symmetry energy.

Recently it had been discussed that future observations of massive
neutron stars may constrain the maximum sound velocity as well as
the coupling parameter in the scalar-tensor theory of
gravity\,\cite{Sotani:2017pfj}. While we confirm this result, we
would also like to emphasize that once the equation of state is
constrained using canonical- and low-mass neutron star observations,
and after a consensus has been reached on the upper limit of the
neutron-star mass, only then one can place useful constraints on the
coupling parameters of the scalar-tensor gravity. We would like to
note that our work does not take into account the possibility that
the core of neutron stars may have exotic degrees of freedom, such
as hyperons, meson condensates, or quarks. In particular, the
central density in neutron stars with $M \gtrsim 1.4
M_{\odot}$---for which spontaneous scalarization occurs---reaches
nuclear densities of $\rho_{\rm c} \gtrsim 2.5 \rho_0$, where our
knowledge of strong interaction is limited. Certainly there could
still be a degeneracy between the scalar-tensor model of gravity and
models of strong interaction at high densities. Much collaborative
theoretical and observational efforts of both nuclear physics and
gravitational physics community are therefore required on this
front.

%%%%%%%%%%%%%%%%%%%%%%%%%%%%%%%%%%%%%%%%%%%%%%%%%%%%%%%%%%%%%%%%%

\section*{Acknowledgments} \vspace{-0.4cm} The author is grateful to
Jorge Piekarewicz for providing the RPA code to calculate the
crust-core transition properties, and to Xiao-Tao He, Charles J.
Horowitz, Bao-An Li, and William G. Newton for many fruitful
discussions. This work is supported in part by the U.S. Department
of Energy Office of Science, Office of Nuclear Physics under Awards
DE-FG02-87ER40365 and DE-SC0018083 (NUCLEI SciDAC-4 Collaboration)
and by the Summer Grant from the Office of the Executive Vice
President and Provost of Manhattan College.
%\end{acknowledgments}

\end{document}